\pdfoutput=1
\documentclass[a4paper,american,citeautoscript,floatfix,longbibliography,pdftex,superscriptaddress,%
aip,jcp,%
reprint,twocolumn,final% add final to see real layout (e.g., no todos anymore)
]{revtex4-1}
\usepackage{amsfonts,amsmath,amssymb}
\usepackage{dcolumn}
\usepackage[T1]{fontenc}
\usepackage[utf8]{inputenc}
\usepackage{graphicx}
\usepackage{microtype}
\usepackage[textsize=footnotesize]{todonotes}
\usepackage{hyperref,hypernat}
\usepackage[todos]{CMImacros}

\newcolumntype{d}[1]{D{.}{.}{#1}}

\newcommand{\trpcage}{Trp~cage\xspace}

\newcommand{\cfeldesy}{\affiliation{Center for Free-Electron Laser Science, Deutsches
      Elektronen-Synchrotron DESY, Notkestrasse 85, 22607 Hamburg, Germany}}%
\newcommand{\zewailcity}{\affiliation{University of Science and Technology, Zewail City, 6th of
      October City, Giza, Egypt}}%
\newcommand{\uhhcui}{\affiliation{The Hamburg Center for Ultrafast Imaging, Universität Hamburg,
      Luruper Chaussee 149, 22761 Hamburg, Germany}}%
\newcommand{\uhhphys}{\affiliation{Department of Physics, Universität Hamburg, Luruper Chaussee 149,
      22761 Hamburg, Germany}}%
\newcommand{\uhhchem}{\affiliation{Department of Chemistry, Universität Hamburg,
      Martin-Luther-King-Platz 6, 20146 Hamburg, Germany}}%
\newcommand{\jkemail}{\email[]{jochen.kuepper@cfel.de}}%
\newcommand{\cmiweb}{\homepage{https://www.controlled-molecule-imaging.org}}%

\begin{document}
\title{Robust and accurate computational estimation of the polarizability tensors of macromolecules}
\author{Muhamed Amin}\cfeldesy%
\author{Hebatallah Samy}\zewailcity%
\author{Jochen Küpper}\jkemail\cmiweb\cfeldesy\uhhphys\uhhchem\uhhcui%
\date{\today}%
\begin{abstract}\noindent% 150 word limit
   Alignment of molecules through electric fields minimizes the averaging over orientations, \eg, in
   single-particle-imaging experiments. The response of molecules to external ac electric fields is
   governed by their polarizability tensor, which is usually calculated using quantum-chemistry
   methods. These methods are not feasible for large molecules. Here, we calculate the
   polarizability tensor of proteins using a regression model that correlates the polarizabilities
   of the 20 amino acids with perfect conductors of the same shape. The dielectric constant of the
   molecules could be estimated from the slope of the regression line based on Clausius–Mossotti
   equation. We benchmark our predictions against the quantum-chemistry results for the \trpcage
   mini protein and the measured dielectric constants of larger proteins. Our method has
   applications in computing laser-alignment of macromolecules, for instance, benefiting single
   particle imaging, as well as for the estimation of the optical and electrostatic characteristics
   of proteins and other macromolecules.
\end{abstract}
\maketitle

\section*{TOC Graphic}
\label{sec:toc-graphic}
\begin{center}
   \includegraphics[height=50mm]{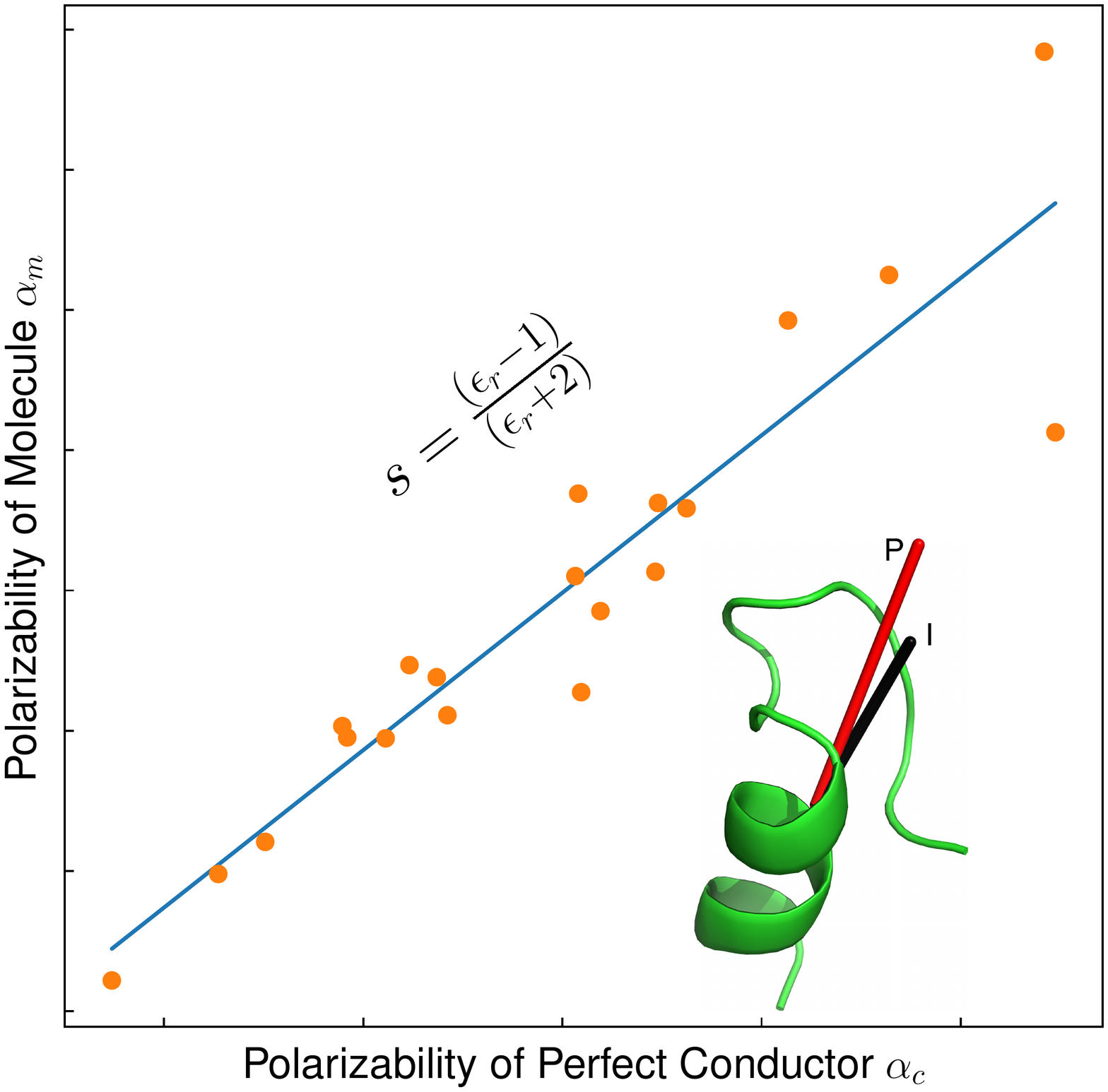}%
\end{center}
\hrulefill\\

\noindent%
One of the main challenges in single particle imaging is recovering the orientation of the imaged
particle from the sparse data in an individual diffraction pattern. It was proposed to use
sophisticated computer algorithms to classify patterns accordingly, but these are highly dependent
on the amount and quality of the available data~\cite{Ayyer:JACR49:1320}. Alternatively, the
molecules can be aligned or oriented by an external field before they are
imaged~\cite{Spence:PRL92:198102, Filsinger:PCCP13:2076, Barty:ARPC64:415}. For small molecules,
this was demonstrated to enable experimental averaging of molecular-frame diffraction signals from
hundred-thousands of shots~\cite{Hensley:PRL109:133202, Kuepper:PRL112:083002, Stern:FD171:393},
whereas for large macromolecules currently achievable degrees of alignment and
orientation\cite{Holmegaard:PRL102:023001, Nevo:PCCP11:9912, Chang:IRPC34:557,
   Karamatskos:arXiv1807:01034} would enable a strong reduction of the phase-space volume for
orientational classification.

Molecules in the gas phase can be aligned and trapped in Stark-effect potentials using intense
non-resonant laser pulses~\cite{Friedrich:PRL74:4623, Stapelfeldt:RMP75:543}. At high frequency,
this effect is dominated by the interactions between the induced dipole moment of the molecules and
the electric field of the laser pulses. These interactions are characterized mainly by the
molecules' static polarizability and its anisotropy, which are described by the polarizability
tensor of the molecule.

The molecular polarizability $\alpha$ is directly related to its electronic properties. It was shown
that the polarizability is directly proportional to a molecule's volume and inversely proportional
to its ionization energy~\cite{Brinck:JCP98:4305}. Furthermore, the value of $\alpha$ is related to
the dielectric constant of the molecule by the Clausius–Mossotti relation and to its refractive
index by the Lorentz–Lorenz equation~\cite{Jansen:PR112:434}. For small molecules the correlation
between the polarizability and molecular volume, ionization energy, electronegativity, and hardness
was investigated extensively using \emph{ab-initio} calculations and density functional theory
(DFT)~\cite{Brinck:JCP98:4305, Blair:JPC141:074306, Hohm:JPCA116:697}.

\begin{table*}
   % put table here so it is typeset a page earlier and everything is arranged more conveniently
   \centering%
   \scriptsize
   \begin{tabular}{lrrr@{\hspace{2em}}d{3.2}d{3.2}d{3.2}@{\hspace{2em}}rrr@{\hspace{2em}}rrr@{\hspace{2em}}d{4.2}d{4.2}d{4.2}@{\hspace{2em}}ccr}
     \hline\hline
     & \multicolumn{3}{c}{$\alpha~(\text{\AA}^3)$} % to show you how to do this "right" ;-)
     & \multicolumn{3}{c}{$k$}
     & \multicolumn{3}{c}{$\alpha_\parallel~(\text{\AA}^3)$}
     & \multicolumn{3}{c}{$\alpha_{\perp}~(\text{\AA}^3)$}
     & \multicolumn{3}{c}{$\Delta{\alpha}~(\text{\AA}^3)$}
     & \multicolumn{3}{c}{$\theta~(\degree{})$} \\[1pt]
     \hline
     & \multicolumn{1}{r}{$\alpha_c$} & \multicolumn{1}{r}{$\alpha_m$} & \multicolumn{1}{c}{$\alpha_p$}
     & \multicolumn{1}{c}{$k_c$}  & \multicolumn{1}{c}{$k_m$} & \multicolumn{1}{c}{$k_p$}
     & \multicolumn{1}{r}{$\alpha_{\parallel{}c}$}  & \multicolumn{1}{r}{$\alpha_{\parallel{}m}$} & \multicolumn{1}{c}{$\alpha_{\parallel{}p}$}
     & \multicolumn{1}{r}{$\alpha_{\perp{}c}$}  & \multicolumn{1}{r}{$\alpha_{\perp{}m}$} & \multicolumn{1}{c}{$\alpha_{\perp{}p}$}
     & \multicolumn{1}{r}{$\Delta{\alpha{}c}$}  & \multicolumn{1}{r}{$\Delta{\alpha{}m}$} & \multicolumn{1}{c}{$\Delta{\alpha{}p}$}
     & $\theta_m$ & $\theta_c$ & \multicolumn{1}{r}{$\theta_{pa}$} \\[1pt]
     \hline
     Gly & 17 & 5 & 8 & 0.19 & 0.11 & 0.09 & 22 & 6 & 9 & 14 & 5 & 8 & 8.25 & 1.23 & 2.33 & 10 & 38 & 48 \\
     Ala & 22 & 7 & 9 & 0.15 & 0.09 & 0.07 & 28 & 8 & 10 & 20 & 6 & 9 & 7.95 & 1.62 & 2.31 & 18 & 5 & 23 \\
     Ser & 25 & 8 & 10 & 0.1 & 0.07 & 0.05 & 29 & 8 & 10 & 22 & 7 & 10 & 6.53 & 1.38 & 2.2 & 36 & 37 & 28 \\
     Pro & 29 & 9 & 11 & 0.12 & 0.06 & 0.06 & 35 & 10 & 11 & 26 & 9 & 11 & 9.29 & 1.1 & 2.41 & 41 & 28 & 13 \\
     Val & 33 & 10 & 12 & 0.19 & 0.1 & 0.09 & 44 & 12 & 13 & 28 & 10 & 12 & 15.99 & 2.68 & 2.92 & 5 & 3 & 5 \\
     Thr & 31 & 9 & 11 & 0.2 & 0.1 & 0.09 & 41 & 11 & 13 & 26 & 9 & 11 & 15.37 & 2.49 & 2.87 & 24 & 20 & 7 \\
     Cys & 28 & 10 & 11 & 0.2 & 0.12 & 0.09 & 38 & 12 & 12 & 23 & 8 & 10 & 14.89 & 3.37 & 2.84 & 30 & 27 & 6 \\
     Ile & 40 & 12 & 14 & 0.24 & 0.1 & 0.11 & 58 & 15 & 17 & 31 & 11 & 13 & 27.03 & 3.39 & 3.75 & 6 & 10 & 8 \\
     Leu & 40 & 10 & 14 & 0.19 & 0.06 & 0.09 & 56 & 11 & 16 & 33 & 10 & 13 & 23.38 & 1.62 & 3.48 & 34 & 15 & 22 \\
     Asn & 34 & 10 & 12 & 0.24 & 0.09 & 0.11 & 50 & 11 & 15 & 26 & 9 & 11 & 24.07 & 2.58 & 3.53 & 38 & 36 & 2 \\
     Asp & 32 & 11 & 12 & 0.22 & 0.09 & 0.1 & 45 & 12 & 14 & 25 & 10 & 11 & 19.69 & 2.67 & 3.2 & 41 & 40 & 1 \\
     Gln & 41 & 12 & 14 & 0.3 & 0.12 & 0.14 & 66 & 14 & 18 & 29 & 10 & 12 & 36.46 & 3.82 & 4.47 & 24 & 31 & 10 \\
     Lys & 44 & 12 & 15 & 0.22 & 0.09 & 0.11 & 64 & 15 & 18 & 34 & 11 & 14 & 29.69 & 3.31 & 3.96 & 8 & 24 & 15 \\
     Glu & 40 & 14 & 14 & 0.25 & 0.13 & 0.12 & 61 & 17 & 17 & 30 & 12 & 12 & 30.63 & 5.11 & 4.03 & 29 & 38 & 9 \\
     Met & 46 & 13 & 15 & 0.37 & 0.15 & 0.17 & 80 & 17 & 21 & 29 & 11 & 12 & 51.16 & 5.93 & 5.58 & 39 & 38 & 3 \\
     His & 44 & 14 & 15 & 0.31 & 0.15 & 0.15 & 72 & 18 & 20 & 30 & 12 & 12 & 41.92 & 6.1 & 4.88 & 31 & 32 & 1 \\
     Phe & 51 & 17 & 17 & 0.33 & 0.18 & 0.15 & 84 & 22 & 22 & 34 & 14 & 14 & 49.44 & 8.13 & 5.45 & 6 & 6 & 4 \\
     Arg & 64 & 15 & 20 & 0.4 & 0.16 & 0.18 & 115 & 20 & 29 & 39 & 12 & 15 & 76.2 & 7.04 & 7.48 & 28 & 30 & 3 \\
     Tyr & 56 & 18 & 18 & 0.36 & 0.19 & 0.17 & 96 & 24 & 25 & 36 & 14 & 14 & 60.24 & 9.62 & 6.27 & 4 & 4 & 4 \\
     Trp & 64 & 22 & 20 & 0.32 & 0.19 & 0.15 & 103 & 28 & 26 & 44 & 18 & 16 & 59.06 & 10.18 & 6.18 & 2 & 19 & 17 \\
     C$_{60}$ & 129 & 78 & 37 & 0.0 & 0.0 & 0.01 & 129 & 78 & 32 & 129 & 78 & 41 & 0.0 & 0.0 & 1.71 & - & - & - \\
     \trpcage\hspace{1ex} & 837 & 216 & 220 & 0.16 & 0.04 & 0.08 & 1048 & 232 & 235 & 732 & 208 & 213 & 315.53 & 24.09 & 25.6 & 15 & 7 & 8 \\
     PSII & 3.6e5 & --  & 1.0e5 & 0.23 & \text{--}  & 0.11 & 5.2e5 & -- & 1.0e5 & 2.8e5 & -- & 1.2e5 & 2.4e5 & \text{--} & 3.0e4 & -- & 37 & -- \\
     \hline\hline
   \end{tabular}
   \caption{Polarizability parameters for the full set of molecules from \emph{ab initio}
      calculations and the model developed in this work. The molecules are sorted according to their
      molecular weight. $\alpha$, $k$, $\alpha_\parallel$, $\alpha_\perp$, and $\Delta\alpha$ are
      defined according to \eqref{eqn:polarizability}--\eqref{eqn:polarizability:delta},
      respectively. The subscripts $m,c,p$ depicts parameters calculated using DFT, calculated using
      ZENO, and predicted based on the regression model, respectively. $\theta_m$ and $\theta_c$ are
      the angles between the principle axis of inertia and the most polarizable axis calculated with
      DFT and ZENO, respectively. $\theta_{pa}$ is the angle between the most polarizable axes as
      calculated with DFT and ZENO. C$_{60}$ was explicitely included as a case not covered by our
      basis set, see text for details. No angles are reported for C$_{60}$, which has icosahedral
      symmetry and is a spherical top. DFT values for photosystem II (PS~II) were not calculated.}
   \label{tab:polarizability}
\end{table*}

In principle, the polarizability tensor can easily be calculated using standard quantum-chemistry
packages. However, to avoid these often expensive calculations, semiempirical methods have been
employed to calculate molecular polarizabilities from atomic polarizabilities. These calculations
show that the orientation of the anisotropic atomic polarizabilities lies along the bond's direction
and it obeys a distance-dependent function for the polarizabilities in the direction of the unbound
atoms~\cite{Miller:JACS112:8543}. Another empirical study showed that the average molecular
polarizability depends on the hybridization of the atoms' orbitals, but not on the type of
atoms~\cite{Miller:JACS101:7213}, \ie, the same atom would have different contributions to the
molecular polarizability depending on its coordination.

Currently, calculating the molecular-polarizability tensor(s) for large molecules is challenging, as
it requires expensive computational resources. Here, we provide a fast and reliable method for
calculating the static-polarizability tensor of macromolecules. First, the polarizability tensor of
a perfect conductor of the molecule's shape is calculated by solving Laplace's equation with
Dirichlet boundary conditions and using Monte-Carlo-path-integral
methods~\cite{Mansfield:PRA64:061401}. Then, the Clausius–Mossotti relationship is used to relate
these polarizabilities to the corresponding molecular polarizabilities through a linear regression
model. In the current demonstration of this approach, we aim at specifically predicting the
polarizability tensor of biological macromolecules, such as proteins, based on the polarizabilities
of the 20 amino acids specified in \autoref{tab:polarizability}. The model is benchmarked against
the \trpcage mini protein and we compare the resulting dielectric constant $\epsilon_r^\text{comp}$
to the measured $\epsilon_r^\text{exp}$ of larger proteins~\cite{Kukic:JACS135:16968}. Furthermore,
we provide a prediction of the polarizability tensor of the prototypical large protein complex
photosystem~II.

Based on the principal moments of polarizability $\alpha_{ii}$, $i=1,2,3$, the average molecular
polarizability $\alpha$ and the polarizability anisotropy $k$ are defined as following:
\begin{align}
  \alpha &=\frac{\alpha_{11}+\alpha_{22}+\alpha_{33}}{3} \label{eqn:polarizability} \\
  k &=\sqrt{\frac{[(\alpha_{11}-\alpha)^2+(\alpha_{22}-\alpha)^2+(\alpha_{33}-\alpha)^2]}{6\alpha^2}} \label{eqn:polarizability:k}
\end{align}
In addition, it is instructive to calculate $\alpha_{\parallel}$ and $\alpha_{\perp}$ to
``visualize'' the polarizability ellipsoid as well as $\Delta{\alpha}$, which is the relevant
quantity for the laser alignment of the most-polarizable axis (MPA)~\cite{Friedrich:JPC99:15686}:
\begin{align}
  \alpha_\parallel &= \max\left(\alpha_{11},\alpha_{22},\alpha_{33}\right) \label{eqn:polarizability:para} \\
  \alpha_\perp     &= \left(3\alpha-\alpha_\parallel\right)/2 \label{eqn:polarizability:perp} \\
  \Delta{\alpha}   &= \alpha_\parallel-\alpha_\perp \label{eqn:polarizability:delta}
\end{align}
Furthermore, to identify the orientation of the molecules' structures with respect to their
polarizability frame, we calculated the angle $\theta$ between the MPA and the molecular $a$ axis,
\ie, the principal axis of inertia with the smallest moment of inertia, \ie, largest rotational
constant, see \autoref{tab:polarizability}.

To build the regression model, the polarizability tensors of 20 amino acids were calculated using
standard quantum-chemistry approaches, \ie, density-functional theory (DFT) starting from
single-conformer structures obtained from New York University's MathMol
database~\cite{NYU:MathMol:2009}, using the 6-31G($d,p$) basis set, the B3LYP functional, and the
Gaussian 09~\cite{Gaussian:2009A02} software package. Then, 20 perfect conductors of the shapes of
the amino acids were constructed by spheres corresponding to the respective van-der-Waals radii of
the constituent atoms. The software ZENO~\cite{Juba:JRNIST122:20} was used to calculate the
polarizability tensor for these perfect conductors by solving Laplace equation using Monte Carlo
numerical-path integration~\cite{Mansfield:PRA64:061401}. We used 1~million Monte Carlo steps for
all calculations. We tested the effect of increasing the number of steps for tryptophan and alanine,
using 100~million steps, and no significant differences were observed, \ie, the differences between
the calculated polarizabilities were smaller than $0.5~\%$. The correlation is studied for $\alpha$,
$k$, $\alpha_\parallel$, $\alpha_\perp$ and $\Delta\alpha$. The values of the average
polarizabilities and the polarizability anisotropies for both approaches are tabulated in
\autoref{tab:polarizability}.

The analytical relationship between the molecular polarizability $\alpha_m$ and polarizabilities of
the perfect conductors $\alpha_c$ could be identified through the Clausius–Mossotti relation:
\begin{align}
  \lim_{\epsilon_{r}\rightarrow\infty} \frac{4\pi\alpha_{c}}{3V} &= \frac{\epsilon_{r}-1}{\epsilon_{r}+2}\\
  \Rightarrow\alpha_c &= \frac{3V}{4\pi}
  \label{eqn:alpha_c}
\end{align}
with the conductor's polarizability $\alpha_{c}$, its volume $V$, and its dielectric constant
$\epsilon_r$. For a molecule that has the same shape, but a finite dielectric constant, the
molecular polarizability $\alpha_m$ is:
\begin{equation}
   \label{eqn:alpha_m}
   \frac{4\pi\alpha_{m}}{3V} =\frac{\epsilon_{r}-1}{\epsilon_{r}+2}
\end{equation}
Substituting \eqref{eqn:alpha_c} in \eqref{eqn:alpha_m} yields:
\begin{equation}
   \alpha_{m}=\alpha_c\frac{\epsilon_{r}-1}{\epsilon_{r}+2}
   \label{eqn:alpha:empirical}
\end{equation}

\begin{figure}[t]
   \includegraphics[width=\linewidth]{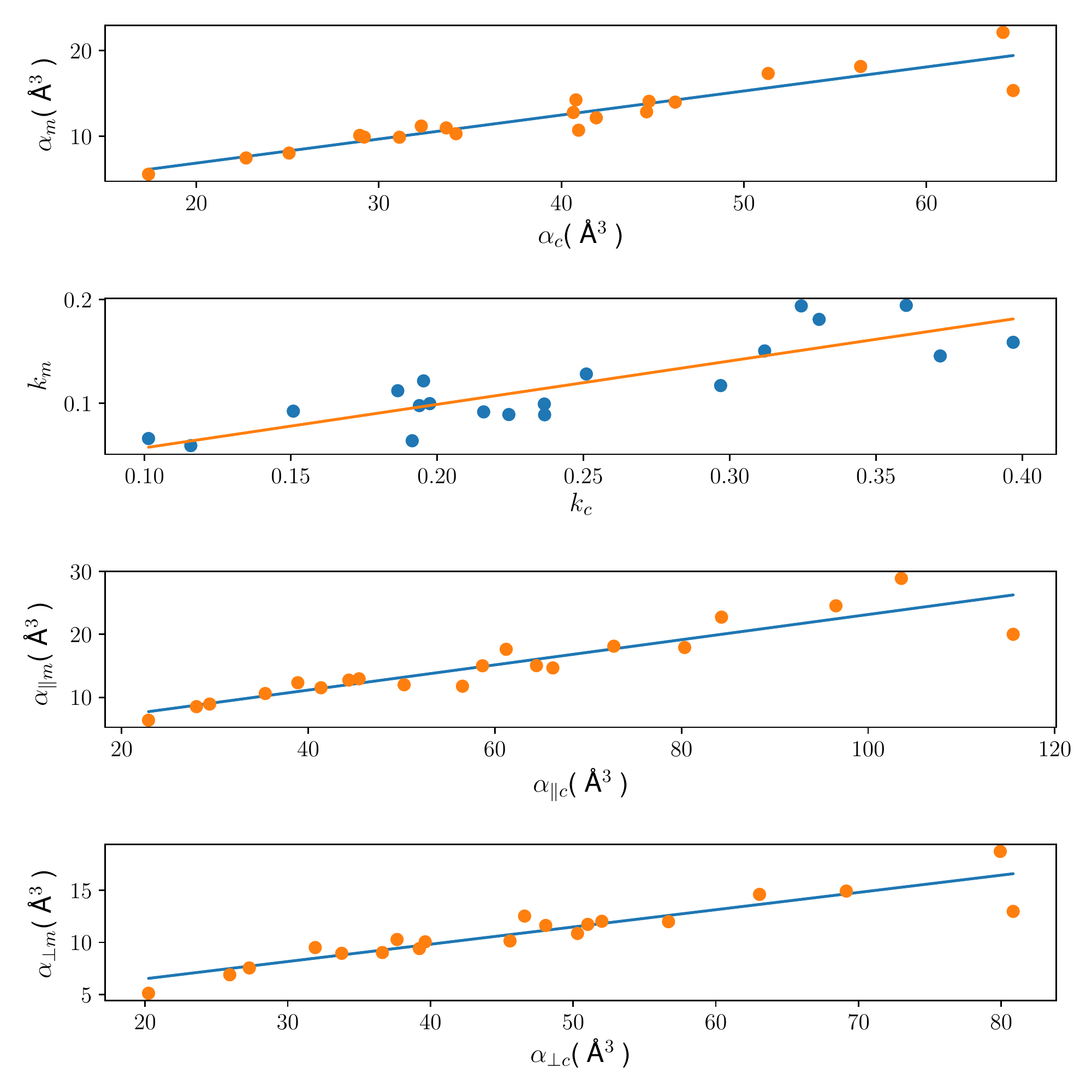}
   \caption{Correlations between a) the average molecular polarizabilities and the polarizabilities
      of same-shape conductors (in \AA$^3$) for the 20 amino acids, b) the polarizability
      anisotropy, c) $\alpha_{\parallel}$, and d) $\alpha_{\perp}$. The raw data are presented in
      \autoref{tab:polarizability}; see text for details, \eg, the parameters obtained from the
      linear regression.}
   \label{fig:correlation}
\end{figure}
The correlations between the molecular polarizability, \eg, computed quantum chemically, and the
polarizability of a perfect conductor of the same shape, calculated with ZENO, for the 20~amino
acids are shown in \autoref{fig:correlation} for $\alpha$, $k$, $\alpha_\parallel$, and
$\alpha_\perp$, respectively. The polarizabilities obtained from the quantum-chemistry calculations
were significantly lower than the values for the perfect conductors, see \autoref{fig:correlation}
and \autoref{tab:polarizability}. However, the polarizability showed a strong correlation between
the values from the two methods, which we analyzed through linear regressions of the individual
pairs $(j_m,j_c)$ for $j=\alpha, k, \alpha_\parallel, \alpha_\perp$. For the average polarizability
$\alpha$ a slope $s_\alpha=0.28$ was obtained with a correlation coefficient $R_\alpha=0.92$ and,
similarly, $s_k=0.42$ with $R_k=0.85$, $s_{\alpha_\parallel}=0.20$ with $R_{\alpha_\parallel}=0.92$,
and $s_{\alpha_\perp}=0.42$ with $R_{\alpha_\perp}=0.93$. We crosschecked our model against the
larger \mbox{6-311G($3df,3pd$)} basis set for all amino acids and obtained practically the same
correlation. According to \eqref{eqn:alpha:empirical} the slope
$s_\alpha=(\epsilon_r-1)/(\epsilon_r+2)$ yields a dielectric constant $\epsilon_{r,\text{p}}=2.17$
averaged over the 20 amino acids.

Using the regression model that includes only the amino acids to predict the molecular
polarizability of the \trpcage mini protein, using the PDB structure 1L2Y~\cite{Neidigh:NSMB:7253},
we predict a value of $\alpha_p=234$, which is 8~\% larger than the value $\alpha_m=216$ from the
quantum-chemistry calculation. Considering that this prediction takes a few seconds, in a
single-core calculation, whereas the DFT quantum-chemistry calculations take more than 48~h on
24~cores on the same computer, this good agreement is extremely satisfying.

Adding the \trpcage data to the regression model of the average polarizability $\alpha$ the
correlation coefficient increases to $R_\alpha=0.998$, the slope decreases to $s_\alpha=0.26$ and
the resulting dielectric constant decreases to \mbox{$\epsilon_{r,\text{p}}=2.05$}. The predicted
polarizabilities $\alpha_\text{p}$ and anisotropy components
$k_\text{p}, \alpha_{\parallel\text{p}}, \alpha_{\perp\text{p}}$ based on the regression model are
given in \autoref{tab:polarizability}.

To analyze the limits of our model, we used it to predicted the properties of C$_{60}$. The
predicted \mbox{$\epsilon_{r,\text{p}}=5.1$} is in fair agreement with the experimental value
\mbox{$\epsilon_{r,\text{exp}}=4.4(2)$}~\cite{Hebard:APL59:2109}. However, there are large
differences between the predicted $\alpha$, $\alpha_\parallel$ and $\alpha_\perp$ and the
quantum-chemistry values for $C_{60}$. This can be attributed to the fact that our model basis does
not contain similar molecules in terms of shape and composition and this comparison is instructive
in setting the limitations of such a basis-set based model, \ie, it is clear that a different basis
set is required to predict the properties of fullerenes.

\begin{table}
   \centering
   \begin{tabular}{l@{\hspace{1em}}c@{\hspace{1em}}c}
     \hline\hline
     protein & $\epsilon_{r,\text{exp}}$ & $\epsilon_{r,\text{p}}$ \\
     \hline
     ACBP   & 3.5 / 3.0 & 2.9 \\
     Av.Pc  & 2.5 / 3.0 & 3.3 \\
     P.1 Pc & 2.0 / 2.5 & 2.9 \\
     hGRx   & 2.0 / 2.0 & 2.3 \\
     \hline\hline
   \end{tabular}
   \caption{Dielectric constants of different proteins from ref.~\onlinecite{Kukic:JACS135:16968},
      see text for specification, PDB numbers, and details. The two experimental values specify
      results from alternative analyses~[\onlinecite{Kukic:JACS135:16968}]. Predicted values
      $\epsilon_{r,\text{p}}$ were obtained through \eqref{eqn:alpha:empirical}.}
   \label{tab:predictions}
\end{table}
For large proteins quantum-chemistry calculations of the polarizabilities are not feasible. Thus, we
compare predictions from our model to experimental dielectric constant, see
\autoref{tab:predictions}. We have performed this comparison for the Acyl-CoA binding protein
(ACBP), Plastocyanin from Anabaena variabilis (Av.Pc), Plastocyanin from Phormidium laminosum (P.1
Pc), and Human Glutaredoxin(hGRx); these structures were obtained from the protein data bank with
PDB codes 1HB6~\cite{VanAalten:JMB:181}, 2GIM~\cite{Schmidt:ACD:1022}, 2Q5B~\cite{Fromme:2007}, and
1JHB~\cite{Sun:JMB280:687}, respectively. The experimental values of $\epsilon_{r,\text{exp}}$ were
obtained by solving the Poisson-Boltzmann equation (PBE) for an ensemble of crystal structures with
the dielectric constant that reproduce the measured $pK_a$ values of a set of amino
acids~\cite{Kukic:JACS135:16968}. The retrieved $\epsilon_r$ of these proteins varies with the
number of structures used in solving the PBE and with the number of observed or absent
NMR-chemical-shift perturbations (CSPs) associated with each ionizable group. In context, our
predicted values are in a good agreement with these experiment-based values. This prediction of
proper dielectric constants $\epsilon_r$ of proteins is essential for the understanding of
electrostatic interactions inside proteins, which has substantial effects on the calculations of
their physicochemical properties such as the refractive indices $n$ as well as their $pK_a$s and
midpoint potentials $E_m$~\cite{Gilson:JMB:503}. Thus, we point out that the accuracy of such
protein properties could be explored using estimated dielectric constant according to the approach
we present here, instead of using a constant value for all proteins~\cite{Ullmann:JBiolChem394:611}.

Regarding laser alignment, \eg, for molecular-frame single-particle imaging, the molecular
polarizability anisotropy defines the rotational dynamics in external electric fields. We have
calculated the polarizability anisotropies $k$ and $\Delta\alpha$ as well as the parallel
$\alpha_{\parallel}$ and perpendicular components $\alpha_{\perp}$ of the polarizability according
to equations \eqref{eqn:polarizability:k}--\eqref{eqn:polarizability:perp}. The resulting values are
given in \autoref{tab:polarizability} and \autoref{fig:correlation}. Also for these properties there
are strong correlations between the quantum-chemically calculated molecular properties and the
values for a perfect conductor of the same shape. Utilizing the slopes determined from a linear
regression we predict the values $k_p$, $\alpha_{\parallel_p}$, $\alpha_{\perp_p}$, and
$\Delta\alpha$, see \autoref{tab:polarizability}. In comparison to quantum-chemistry values we
obtained standard deviations of 0.02~{\AA}$^3$, 2.3~{\AA}$^3$, 1.1~{\AA}$^3$, and 2.7~{\AA}$^3$,
respectively for the set of the 20 amino acids and \trpcage, which reflects the good agreement of
our model to the DFT calculations. Furthermore, we checked the agreement of the angles between the
MPA and the molecules' inertial $a$ axis, see \autoref{tab:polarizability}, which also has a
significant influence on the rotational dynamics in the laser alignment of complex
molecules~\cite{Hansen:JCP139:234313}. The average of the angles between the predicted and the DFT
MPA is \degree{13}, with a standard deviation of \degree{14}. In general, the largest deviations are
observed for small amino acids, which we ascribe to their highly anisotropic distribution of atoms
around the axes of inertia. However, for large molecules such as \trpcage this effect reduces
significantly. Thus, for macromolecules, the polarizability tensors of the perfect conductors have
orientations in the molecules' inertial frame that are fairly similar to the quantum-chemically
calculated once. This indicates that the calculated rotational/alignment dynamics using values
predicted from our model will reflect the actual molecular dynamics well. We have started such
calculations to predict achievable degrees of alignment for large macromolecules. These should
reliably predict the achievable degrees of alignment of large and complex macromolecules and could
be exploited for angular deconvolution of diffractive-imaging data.

In conclusion, we devised a simple model for the prediction of the static-polarizability tensors of
large or complex molecules based on extremely fast and robust calculations of the polarizability
tensor of the corresponding conductor of the same shape. We benchmarked this model for proteins as
prototypical biological macromolecules using a basis set of 20 amino acids. Furthermore, using the
Clausius-Mosotti equation these results were extended to predictions of the macromolecules'
dielectric constants. The accuracy of the predicted polarizabilities and anisotropies, compared to
values from standard quantum-chemistry calculations, is better than 10~\% and the predicted
dielectric constants are within the error bounds of the experimental values.

These results have important applications in the computational prediction and the quantitative
understanding of laser alignment of macromolecules, \eg, in single-molecule diffractive imaging
experiments, as well as for calculations of the $pK_a$ and $E_m$ values of proteins.

\section*{Acknowledgment}
This work has been supported by the European Research Council under the European Union's Seventh
Framework Programme (FP7/2007-2013) through the Consolidator Grant COMOTION (ERC-614507-Küpper) and
by the Deutsche Forschungsgemeinschaft through the Clusters of Excellence ``Center for Ultrafast
Imaging'' (CUI, EXC~1074, ID~194651731) and ``Advanced Imaging of Matter'' (AIM, EXC~2056,
ID~390715994) of (DFG) and the priority program ``Quantum Dynamics in Tailored Intense Fields''
(QUTIF, SPP~1840, KU~1527/3).

\bibliography{string,cmi}%
\onecolumngrid
\end{document}